\documentclass[12pt]{article}
\usepackage{graphicx}

\newcommand\pubnumber{ }
\newcommand\pubdate{\today}

\def \babar {{\it B{\scriptsize A}B{\scriptsize AR}\ }}
\def \etal{{\it et al.}}
 
\def\Title#1{\begin{center} {\Large #1 } \end{center}}
\def\Author#1{\begin{center}{ \sc #1} \end{center}}
\def\Address#1{\begin{center}{ \it #1} \end{center}}

\newcommand\pubblock{\rightline{\begin{tabular}{l} \pubnumber\\
         \pubdate  \end{tabular}}}
\newenvironment{Abstract}{\begin{quotation}  }{\end{quotation}}
\newenvironment{Presented}{\begin{quotation} \begin{center} 
             PRESENTED AT\end{center}\bigskip 
      \begin{center}\begin{large}}{\end{large}\end{center} \end{quotation}}

\begin{document}
\begin{titlepage}
\pubblock

\vfill
\Title{	Experimental status of $B \to \tau \nu$ and $B \to \ell \nu (\gamma)$}

\vfill
\Author{ Roger Barlow\\(\babar collaboration)}
\Address{The University of Huddersfield, Queensgate, Huddersfield HD1 3DH,UK}
\vfill
\begin{Abstract}
The experimental results of the \babar\ and Belle collaborations are presented on the measurement of the branching ratio
for the decay $B^\pm \to \tau^\pm \nu$, and the limits set on the branching ratios for the decays
$B^\pm \to e^\pm \nu, B^\pm \to e^\pm \nu \gamma,B^\pm \to \mu^\pm \nu$ and $B^\pm \to \mu^\pm \nu \gamma$ 
\end{Abstract}
\vfill
\begin{Presented}
CKM2010, the 6th International Workshop on the CKM Unitarity Triangle, University of Warwick, UK, 6-10 September 2010
\end{Presented}
\vfill
\end{titlepage}
\def\thefootnote{\fnsymbol{footnote}}
\setcounter{footnote}{0}

\section{Introduction: Theory and Experiment}

\begin{figure}[htb]
\centering
\includegraphics[height=1in]{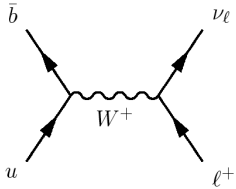}
\caption{Decay of the charged $B$ in the Standard Model}
\label{fig:figone}
\end{figure}

Within the Standard Model the charged $B$ meson can decay through a virtual $W$ to a charged lepton 
and its corresponding neutrino, as shown in Figure~\ref{fig:figone},
and the branching ratio for the process  is given by
\begin{equation}
Br(B \to \ell \nu) = {G_F^2 m_B \over 8 \pi} m_\ell^2 \left(1 - {m_\ell^2\over m_B^2} \right) f_B^2 |V_{ub}|^2 \tau_B
\end{equation}
where $m_B$ and $m_l$ are the masses of the $B$ and of the lepton, $\tau_B$ is the B lifetime, $G_F$ is the Fermi constant,
$V_{ub}$ is the CKM element and $f_B$ is the B meson form factor denoting the extent to which the meson can be considered as a quark-antiqquark pair. All of these numbers are well known, and for the $B \to \tau \nu$ decay the predicted branching ratio is of order $10^{-4}$.  For muon and electron decays it is much smaller, due to the $m_l^2$ factor, which in turn is due to helicity suppression: the spinless $B$ meson, like the pion, prefers to decay to the heaviest possible charged lepton because balancing the spins of the outgoing leptons requires them to have the same handedness, and the neutrino forces its charged partner into the unfavoured helicity.

\begin{figure}[htb]
\centering
\includegraphics[height=1in]{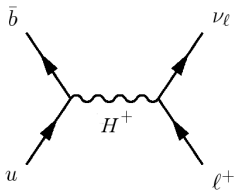}
\caption{Decay of the charged $B$ through a charged Higgs}
\label{fig:figtwo}
\end{figure}

Many ``New Physics"  models contain a charged Higgs boson, which can also mediate the decay, as shown in Figure~\ref{fig:figtwo}.
 A minimal model with an additional Higgs doublet\cite{Hou} predicts a branching ratio from the sum of the two amplitudes
\begin{equation}
Br(B \to \ell \nu) = {G_F^2 m_B \over 8 \pi} m_\ell^2 \left(1 - {m_\ell^2\over m_B^2} \right) f_B^2 |V_{ub}|^2 \tau_B
\times \left( 1-\ tan^2{\beta} {{m_B^2} \over {m_H^2}} \right) ^2 
\end{equation}
where $tan \beta$ is the ratio of the Higgs vacuum expecation values.  A Supersymmetric version \cite{Akeroyd}
predicts
\begin{equation}
Br(B \to \ell \nu) = {G_F^2 m_B \over 8 \pi} m_\ell^2 \left(1 - {m_\ell^2\over m_B^2} \right) f_B^2 |V_{ub}|^2 \tau_B
\times \left( 1-{\ tan^2{\beta}\over 1 + \overline \epsilon_0 \tan \beta} {{m_B^2} \over {m_H^2}} \right) ^2 
\end{equation}
where $\overline\epsilon_0$ is a SUSY correction factor

The $B$ factories have performed several analyses of these decays, using up to 468M  $B \overline B$  pairs at  BaBar and  657M pairs at Belle.  
These are not easy measurements, as there is at least one neutrino in the final state so these $B$ mesons  are not reconstructable. 
A tagging technique is used:  the decay of the $\Upsilon(4S)$ produces $B$ mesons  in pairs: 
if one (the tag)  is reconstructed,  the rest of the event must be a $B$ meson.

There are two classes of tags. In {\em hadronic tags} the charged and neutral particles
are identified, using the excellent $\pi/K$ separation which both BaBar and Belle enjoy, and 
reconstructing $\pi^0$ and $K^0$ mesons. From these the analysis attempts to construct
 heavier mesons such as the $D$, $D^*$ and $J/\psi$;  if successful it then attempts
to combine the object with further light hadrons to reconstruct a $B$ meson. 
The efficiency is low -- of order $10^{-3}$ -- 
but a reasonably pure sample of $B$ mesons is obtained.
In {\em semileptonic tags} a charmed meson, $D$ or $D^*$, is constructed as before.
Then a high momentum lepton ($\mu$ or $e$) is required. The efficiency is higher, of order $10^{-2}$,
though the sample is not so pure.
The two tag methods are very different, and give independent data samples, so the two analyses are generally reported separately.

Details of all the cuts used depend on the analysis. There is also a difference of approach: some analyses find a tag and then look for a  pure leptonic $B$ decay in the rest of the event, others
find the signal  first and then look for the tag.  References should be consulted for complete accounts.

\section{Limits on decays to light leptons}

\subsection {The decays $ B \to e \nu$ and $B \to \mu \nu$ }

For the \babar hadronic tag analysis~\cite{emu} a  high momentum charged  lepton is first searched for, and then the requirement that all other observed particles combine to form a system with the mass of the $B$ meson is applied, whereas for the semileptonic analysis~\cite{babarsemi} the tag is applied first and the lepton requirement second. 
The main search variables used are the energy-substituted mass, $m_{ES}=\sqrt{E_{beam}^2 - |\vec p_B|^2}$, and $P_{FIT}$,  a linear combination of the lepton momentum the the $B$ rest frame and in the centre of mass, the combination being a Fisher discriminant chosen to suppress backgrounds. 

\begin{figure}[htb]
\centering
\includegraphics[height=3.5in]{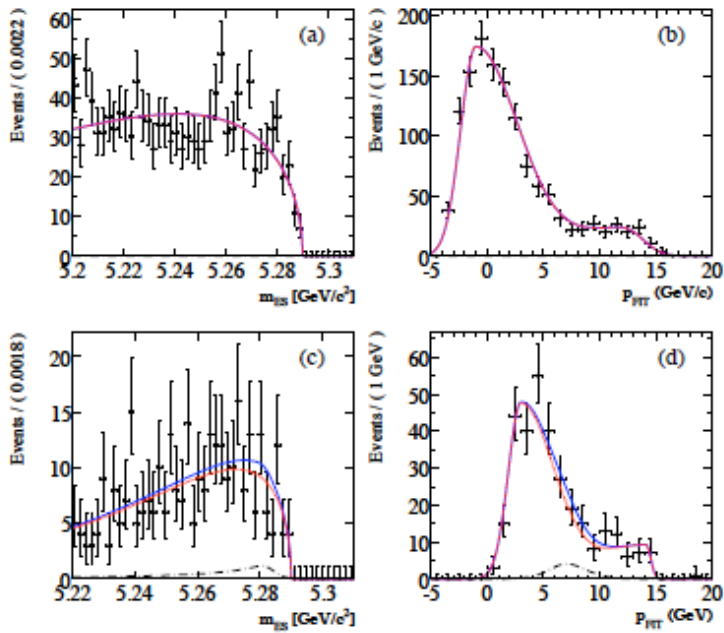}
\caption{Data (crosses) and fits (curves) to the variables $m_{ES}$ (left) and $p_{FIT}$ (right) for the muon (top) and electron (bottom) channels. Taken from \cite{emu}. }
\label{fig:figTHREE}
\end{figure}

The hadronic analysis \cite{emu} turns out to be more powerful than the semileptonic\cite{babarsemi}. 
Figure~\ref{fig:figTHREE} shows the results. There is no sign of an excess at $m_{ES}=5.28$ and high $p_{FIT}$

This gives upper limits for the branching ratios of 
$1.0 \times 10^{-6}$ for the muon and $1.9 \times 10^{-6}$ for the electron channel. (These are 90\% confidence limits obtained using a Bayesian technique with a flat prior on the decay rate.)
The Belle results\cite{emuBELLE} are similar:  $1.7 \times 10^{-6}$ for the muon and $0.98 \times 10^{-6}$ for the electron channel, though different statistical procedures means that the numbers should not be compared directly.
But the conclusion is the same: there is no evidence for any signal.

\subsection {The decays $ B \to e \nu \gamma$ and $B \to \mu \nu \gamma$ }

Radiation of a (spin 1)  photon can evade the helicity suppression factor, at the
cost of an extra factor $\alpha$.
The standard model prediction is of order $10^{-6}$.

The technique used is to find a fully reconstructed hadronic $B$ tag, and then require that there be only one extra charged track, identified as an electron or muon. A high energy photon is also required. From the lepton and the photon one then reconstructs the mass of the neutrino. An event is counted as signal if the mass squared falls below
0.46  ${\rm GeV}^2/c^4$ for the electron channel and 0.41 ${\rm GeV}^2/c^4$ for the muon channel. (The presence of  bremsstrahlung photons from electrons makes the analyses different.)

\babar  \cite{emugamma} see 4 events in the electron channel and 7 in the muon channel, however their expected backgrounds (predominantly from semileptonic decays of the $B^\pm \to \pi^0 \ell^\pm \nu$ and  $B^\pm \to \eta \ell^\pm \nu$, where one of the photons from the hadron decay is lost)
 are 2.7 and 3.4 events respectively, so there is no 
evidence for a signal. A combined 90\% upper limit is quoted as $15.6 \times 10^{-6}$. This is model independent
in that nothing is assumed about the direction of the photon.  If specific assumptions are made
about the vector and axial-vector form factors this gives different results: if one assumes that $f_A=f_V$ the limit is 
reduced to $3.0 \times 10^{-6}$.

\section{ Measurements of $B \to \tau \nu$}

This is the channel of greatest interest, as the branching ratio is large enough that it can now be measured, as opposed to
merely having limits set on it. It is also the most difficult, as the decay of the $\tau$ produces at least one extra neutrino, and possibly two.  But we now have  measurements from both experiments wth both tagging techniques, and four consistent measured values emerge.

A tag $B$, in the hadronic or semileptonic mode, is found. The remaining tracks are considered as potential products from a $\tau$ 
decay. All four analyses consider the $e \nu \nu, \mu \nu \nu$ and $\pi  \nu$ decays, some also consider the $\rho \nu$ and/or
the $\pi^\pm\pi^+ \pi^- \nu$ decay. Several requirements(see~\cite{babarsemi,babartauhad,belletausemi,belletauhad} for full details) are laid on the momenta and missing momenta to reduce background. However the lack of any definite mass combination on which a hard cut can be placed means that this is not enough to isolate a signal.

The necessary extra signature is found by summing the total visible energy in the electromagnetic calorimeter
and removing what can be accounted for by known particles in the event. This surplus -- called $E_{ECL}$ by BELLE and $E_{extra}$ by \babar, is close to zero for genuine signal, but generally larger for background events.

\begin{figure}[htb]
\centering
\includegraphics[height=3.5in]{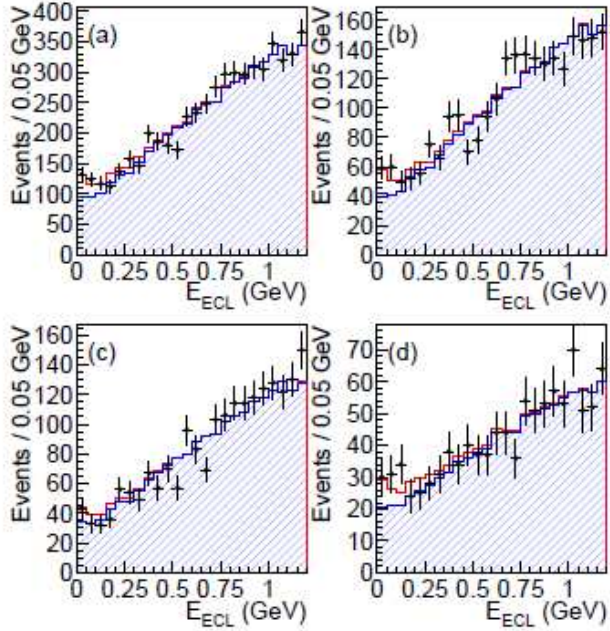}
\caption{The extra anergy for (a) all decay modes (b) $\tau \to e \nu \nu$, (c) $\tau \to \mu \nu \nu$ and (d)
$\tau \to \pi \nu$.  Data points are crosses, blue shading is predicted background, the red curve includes the fitted signal. Taken from Ref.~\cite{belletausemi} }
\label{fig:figfour}
\end{figure}

Fig. \ref{fig:figfour} shows the extra energy distributions from the Belle semileptonic tag analysis. The small (but statistically significant) excess of measured signal over expected background 
in the first few bins of the histogram is the signal being sought.

\begin{figure}[htb]
\centering
\includegraphics[height=3in]{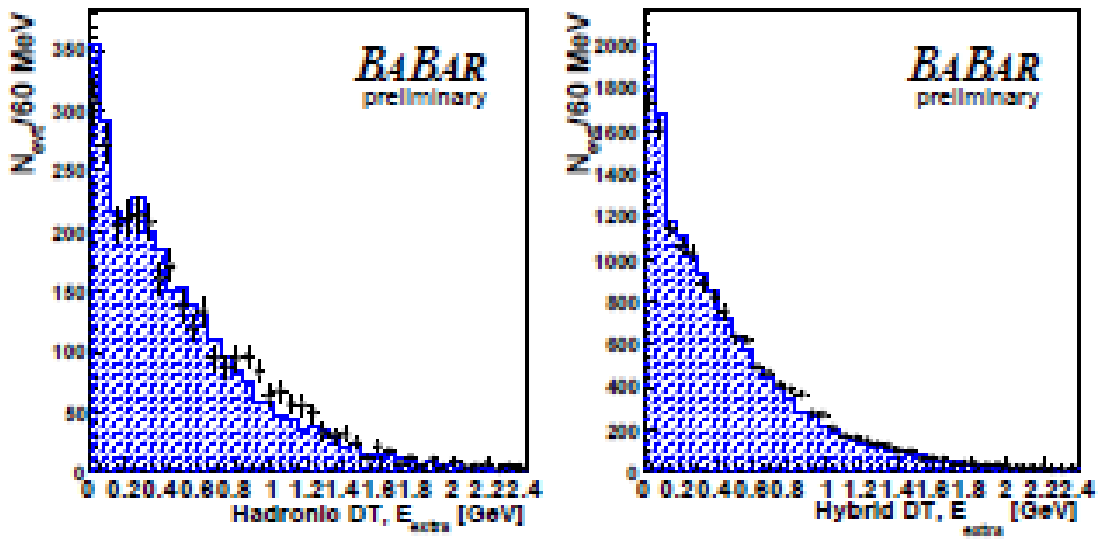}
\caption{The extra energy for double decay modes, the first tag being semileptonic and the second hadronc (left) of semileptonic (right)  Taken from Ref.~\cite{babarICHEP} }
\label{fig:figfive}
\end{figure}

To claim this as an observation of the decay mode $B \to \tau \nu$ requires confidence in the simulation and modelling of the backgrounds. Fortunately this is supplied through events in which
both $B$ meson decays are tagged, at least one semileptonically. The `extra' energy  is shown for such events in Fig.~\ref{fig:figfive} for \babar, displaying
excellent agreement between simulation and data; Belle has similarly impressive plots.

\babar\ quote values of 
$(1.80^{+0.57}_{-0.54} \pm 0.26) \times 10^{-4}$
for the hadronic tag decays~\cite{babarICHEP}  (a new result presented at ICHEP this year) and 
$(1.7\pm 0.87 \pm 0.2) \times 10^{-4}$ for the semileptonic tag, and combine the results (the samples are independent)
to give a combined value of 
$(1.76 \pm 0.49) \times 10^{-4}$.  Belle quote
$(1.79^{+0.56}_{-0.49} {}^{+0.46}_{-0.51}) \times 10^{-4}$ for the hadronic tags and
$(1.54^{+0.38}_{-0.37} {}^{+0.29}_{-0.31}) \times 10^{-4}$ for the semileptonic  tags (also a new result preented at ICHEP).
These values are all compatible, and the Heavy Flavour Averaging Group  has combined them~\cite{HFAG} to give a branching ratio of $(1.64 \pm 0.34) \times 10^{-4}$

\section{Implications}

The leptonic decay $B^\pm \to \tau^\pm \nu$ is thus well established. The branching ratio is in good agreement with the Standard Model 
prediction of $(1.20 \pm 0.25) \times 10^{-4}$, which is evaluated using the HPQCD value for $f_B$ of $190 \pm 13 $ MeV and the
HFAG value for $V_{ub}$ of $ (4.32 \pm 0.16 \pm 0.29) \times 10^{-3}$

\begin{figure}[htb]
\centering
\includegraphics[height=3in,angle=270]{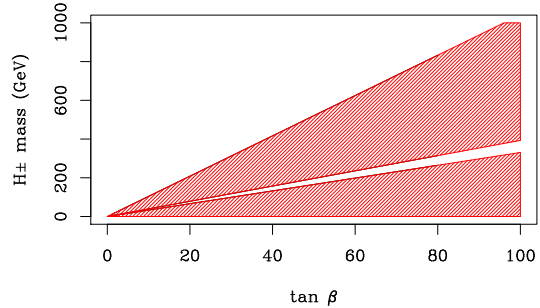}
\caption{Parameter space for Eq. 2 ruled out at the 2 $\sigma$ level }
\label{fig:figsix}
\end{figure}

This good agreement means that the effect of extra processes must be small, and the term $(1-\tan^2 \beta {m_b^2 \over m_H^2})^2$ in Eq. 2,
or its equivalent in Eq. 3, must be close to 1. This means that various parameter values can be ruled out,
in particular a large value of $\tan \beta$ is incompatible with a low mass charged Higgs. This is an important constraint
on model building. Given the ratio $r$ of the measured to the predicted result and the combined error $\sigma$, then
the limit at $n$ standard deviations is given by $M_{H^\pm}> {M_B \tan \beta \over \sqrt {1-(r-n \sigma)}}$, as shown in Fig.\ref{fig:figsix}.
There is a narrow window of possibility,${M_B \tan \beta \over \sqrt {1+(r-n \sigma)}}> M_{H^\pm}> {M_B \tan \beta \over \sqrt {1+(r+n \sigma)}}$ 
but such a fine-tuned coincidence
($\tan^2 \beta m_B^2 \approx 2 m_H^2$) would raise many questions.

As all the terms in the extra part of Eq. 2 are squared, any  BSM effects reduce the branching ratio.   The first preliminary 
measurement from Belle was indeed on the low side, resulting in some excitement. However it was later revised upwards, and the  measurement is now
fully in agreement with the standard model.

However there is another problem associated with this measurement.  If it  is included in an overall global fit to all the relevant data for the CKM triangle, a discrepancy appears. 
The measured value is larger than the preferred one, which is $(0.805 \pm0.071) \times 10^{-4}$ for UTfit~\cite{CKMone} 
and $(0.763^{+0.114}_{-0.061}) \times 10^{-4}$ according to CKMfitter~\cite{CKMtwo}.  The fits adjust the CKM elements and other parameters to fit all the measurements. The form factor $f_B$ may be fit or taken from other calculations, the result is similar. 
That the two  fitting groups agree on this result, though they use different methodologies, is an indication that this should be taken seriously. There appears to be a tension between the value of $V_{ub}$ and the measurements of $sin (2\beta)$.

\section {Conclusions}

While increasingly stringent limits are placed on the decay of the charged $B$ meson to light leptons, the decay $B^\pm \to \tau^\pm \nu$
 is now well established. The results agree with the standard model,  imposing significant constraints on BSM parameter values. 
The measurement is also a source of tension within global fits to the CKM matrix.  The SuperB factory will improve the resolution by an order of magnitude, and this is 
yet another argument for building such a machine.

\end{document}